\documentclass[twocolumn,nopacs,preprintnumbers,amsmath,amssymb]{revtex4}

\usepackage{verbatim}
\usepackage{graphicx}
\usepackage{dcolumn}
\usepackage{bm}

\begin{document}

\title{Comparative study of the catalytic growth of patterned carbon nanotube films}
\author{Christian Klinke}
\email{christian@klinke.org}
\author{Jean-Marc Bonard}
\altaffiliation{Present address: Rolex S.A., 3-7 Rue Francois-Dussaud, 1211 Geneva 24, Switzerland.}
\affiliation{Institut de Physique des Nanostructures, Ecole Polytechnique F\'ed\'erale de Lausanne,\\
CH - 1015 Lausanne, Switzerland}

\author{Klaus Kern}
\affiliation{Institut de Physique des Nanostructures, Ecole Polytechnique F\'ed\'erale de Lausanne,\\
CH - 1015 Lausanne, Switzerland \\ \textnormal{and} \\ Max-Planck-Institut f\"ur Festk\"orperforschung, D - 70569 Stuttgart, Germany}

\begin{abstract}

Three different catalysts (Fe, Ni, Co nitrates dissolved in
ethanol) were patterned on a SiO$_{2}$/Si substrate and multi-wall
carbon nanotubes were grown by catalytic decomposition of
acetylene. We compare the growth of the carbon nanostructures in
the temperature range between 580$^{\circ}$C and 1000$^{\circ}$C.
With our experimental set-up the catalyst solutions of cobalt and
nickel were found to be less efficient than the one of iron. An
optimal production of multi-wall nanotubes was observed at
temperatures between 650$^{\circ}$C and 720$^{\circ}$C with the
iron solution as catalyst. We found a tendency towards thicker
structures with higher temperatures. Finally, we suggest a
mechanism for the growth of these carbon structures.

\end{abstract}

\maketitle

\section{Introduction}

Carbon nanostructures like fullerenes~\cite{SMALLEY},
nanotubes~\cite{IIJIMA}, nano-onions~\cite{UGARTE} and
nano-horns~\cite{TAKAHASHI} have attracted much interest recently.
In particular their mechanical and electronic properties are the
subject of intensive studies~\cite{FORRO,LIEBER}. Beside the
fundamental interest in their physical and chemical properties,
there are already some applications based on carbon nanotubes. For
example, they are capable to work as efficient field
emitters~\cite{CHATELAIN} and can form a basis for very robust
fibers~\cite{POULIN}. Nanotubes can be produced by arc
discharge~\cite{IIJIMA}, by laser-ablation~\cite{SMALLEY2} or by
chemical vapor deposition techniques
(CVD)~\cite{ZHANG,LUCAS,KERN,PROVENCIO,KIM,KIM2,HAN}. CVD is
currently the most promising and flexible method with regard to
applications, but our understanding of the influence of the
catalyst and the deposition parameters on the nanotube growth is
still fragmentary.

We use here the CVD method in combination with microcontact
printing to grow patterned films of multi-wall carbon
nanotubes~\cite{KERN}. Microcontact printing ($\mu$CP) has become
an often applied method in the last few years because it is a
simple way to define chemical patterns on a variety of
substrates~\cite{DELAMARCHE,WHITESIDES,WHITESIDES2}. We use this
method to selectively deliver a catalyst to the substrate surface,
which in turn activates the growth of nanotubes~\cite{KERN}. The
advantage of the patterning is that one can compare the regions
with and without catalyst, and thus exactly determine the role of
the catalyst. In order to better understand the catalytic growth
we systematically examined the parameters temperature, catalyst
composition and concentration.

\section{Experimental methods}

\subsection{Synthesis of nano-structured material}

$<$100$>$-oriented boron doped silicon with the native SiO$_{2}$
oxide layer was used as substrate. The stamps for $\mu$CP were
prepared by curing poly\-(dimethyl)\-siloxane (PDMS) for at least
12~h at 60$^{\circ}$C on a structured master prepared by contact
photolithography. The width of the square patterns on the wafer is
5~$\mu$m. The stamps were subsequently hydrophilized by an oxygen
plasma treatment (O$_{2}$ pressure $\sim $0.8~mbar, load coil
power $\sim $75~W, 60~s). The stamp was loaded with 0.2~ml of
catalyst solution for 30~s and then dried in a nitrogen stream for
10~s.

The solutions were Fe(NO$_{3}$)$_{3}\cdot$9H$_{2}$O,
Ni(NO$_{3}$)$_{2}\cdot$6H$_{2}$O or
Co(NO$_{3}$)$_{2}\cdot$6H$_{2}$O dissolved in ethanol at
concentrations between 50 and 200~mM. The solutions were used 12~h
after preparation because the metallic ions in the solution form
chemical complexes which become larger with time. A period of 12~h
for this ``aging'' of the solution was found to be ideal for the
catalytic growth of nanotubes~\cite{KERN}. The printing was
performed by placing the stamp on the surface of the SiO$_{2}$/Si
wafer for 3~s.

The samples were placed in a horizontal flow reactor (quartz tube
of 14~mm diameter in a horizontal oven) directly after the
printing. The treatment in the CVD oven proceeded in three steps.
In the first step, the catalyst was annealed for 20~min under a
flow of 80 ml/min of nitrogen to roughen the surface of the
catalyst and to clean the reactor atmosphere. The actual
deposition was performed with 80 ml/min of nitrogen plus 20~ml/min
of acetylene (carbon source for the catalytic growth) at
atmospheric pressure for 30~min. The third step was a final
annealing of 10~min. under 80~ml/min of nitrogen. The same
temperature was used throughout the entire procedure, which
implies that a change of the temperature affected all the steps.

\subsection{Characterization techniques}

Scanning electron microscopy (SEM) was performed to analyze the
microstructures in plan view. A Philips XL~30 microscope equipped
with a field emission gun (FEG) operating at an acceleration
voltage between 2 and 5~kV, a working distance of typically 10~mm,
and in secondary electron (SE) image mode was used.

The growth morphology of the tubular structures and their
crystallinity were controlled by transmission electron microscopy
(TEM). For this purpose a Philips EM~430 microscope equipped with
a Gatan image plate operating at 300~kV (point resolution 0.3~nm)
was used.

\section{Results}

We studied systematically the influence of (a) the catalyst
solutions containing iron, nickel or cobalt ions, (b) the
deposition temperature and (c) the concentration of the catalyst
on the nanotube growth. We used concentrations of 50~mM and 100~mM
of the three catalysts at temperatures between 580$^{\circ}$C and
1000$^{\circ}$C in 70$^{\circ}$C steps. For the catalyst iron,
which produced the highest amount of nanotubes, we also examined
samples with a concentration of 150~mM and 200~mM at
650$^{\circ}$C.

\begin{figure}[h]
\begin{center}
\includegraphics[width=0.45\textwidth]{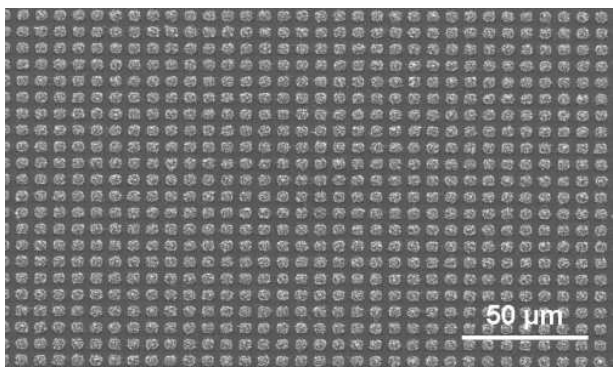}
\caption {\it SEM micrograph demonstrating the homogeneity of the patterning on a Si/SiO$_{2}$ sample (100 mM iron nitrate solution at 720$^{\circ}$C)}
\label{HOMOGENEITY}
\end{center}
\end{figure}

After the catalytic growth of the structures, the observation by
SEM showed a homogeneous growth over the whole sample surface
($>$1~cm$^2$, Fig.~\ref{HOMOGENEITY}). The nanotubes grow only
where the catalyst has been printed, and the bare substrate is
free of any carbon form for growth temperatures below
800$^{\circ}$C.

\begin{figure}[h]
\begin{center}
\includegraphics[width=0.45\textwidth]{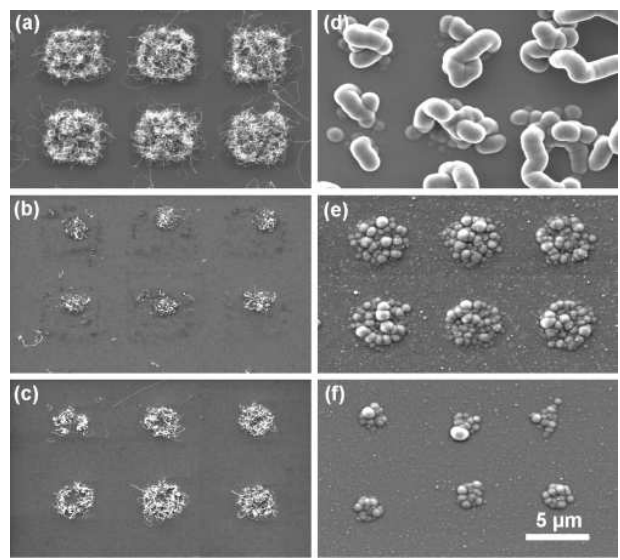} \caption {\it SEM micrographs demonstrating the effect of the catalyst on the nanotube growth: 100 mM solution of (a) Fe (b) Ni (c) Co nitrate at 720$^{\circ}$C and of (d) Fe (e) Ni (f) Co nitrate at 1000$^{\circ}$C} \label{CATALYST}
\end{center}
\end{figure}

As displayed in Fig.~\ref{CATALYST}, we found that the use of the
iron catalyst resulted in the highest fidelity of replication of
the square pattern, whereas the use of a pure nickel or cobalt
solution resulted in inferior patterning. Because of the wetting
behavior on the stamp~\cite{KERN}, the nickel and the cobalt
solutions produced one drop on each square structure of the stamp,
which were transferred to the substrate. As a result, the patterns
obtained with cobalt and nickel are small and of irregular,
circle-like shape. At concentrations higher than 100 mM the
printing of the iron solution became also more difficult. XPS and
TEM experiments have shown that the printed iron catalyst consists
of a gel-like material of partially hydrolyzed Fe(III) nitrate
that forms a porous and continuous Fe$_{2}$O$_{3}$ film after
annealing~\cite{SCHLAPBACH}. We assume that the printed catalysts
nickel and cobalt are of a similar structure.

The comparison of the three catalysts shows that iron produced the
highest density of carbon structures at any considered
temperature, as seen in Fig.~\ref{CATALYST}. Nickel and cobalt
turned out to be not as good catalysts as iron, as we found only
few nanotubes at 720$^{\circ}$C with 100~mM nickel and cobalt
solutions. At 1000$^{\circ}$C, the printed nickel and the cobalt
solution initiated the growth of spherical carbon structures, and
iron caused the growth of thick worm-like structures.

\begin{figure}[h]
\begin{center}
\includegraphics[width=0.45\textwidth]{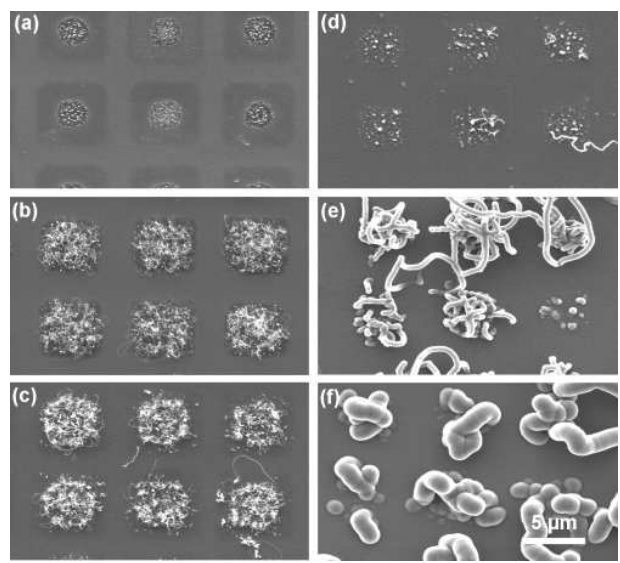} \caption {\it SEM micrographs demonstrating the effect of the temperature on the nanotube growth: 100 mM solution of iron nitrate at (a) 580$^{\circ}$C (b) 650$^{\circ}$C (c) 720$^{\circ}$C (d) 860$^{\circ}$C (e) 930$^{\circ}$C (f) 1000$^{\circ}$C } \label{TEMPERATURE}
\end{center}
\end{figure}

There is a tendency towards thicker structures with increasing
temperature~\cite{RODRIGUEZ}, as shown in Fig.~\ref{TEMPERATURE}
for iron. The catalytic growth of nanotubes started at a
temperature of 620$^{\circ}$C, but we observed only a few
nanotubes at this temperature. A uniform coverage of each printed
square was obtained at temperatures of 650$^{\circ}$C and higher.
The thinnest nanotubes were found at 650$^{\circ}$C with iron as
catalyst. The highest density of nanotubes was observed at
720$^{\circ}$C. Using catalyst concentrations of 100~mM we found
relatively thick worm-like structures at temperatures exceeding
930$^{\circ}$C (``carbon worms''). These structures grew to a
diameter of up to 1~$\mu$m. Furthermore, the acetylene starts to
dissociate in the gas phase at temperatures above 800$^{\circ}$C
and the resultant carbon forms a amorphous continuous layer on the
silicon surface of all samples. This layer gets also thicker with
the temperature.

\begin{figure}[h]
\begin{center}
\includegraphics[width=0.45\textwidth]{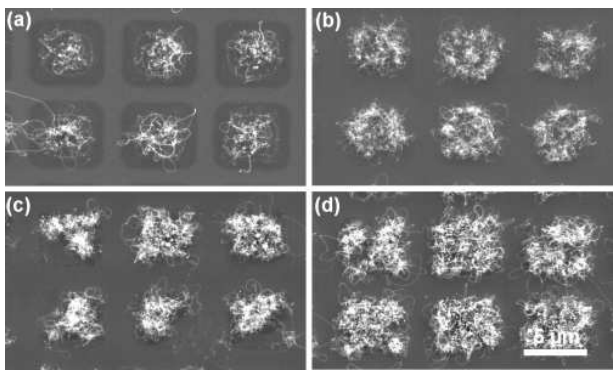} \caption {\it SEM micrographs demonstrating the effect of the concentration on the nanotube growth: a (a) 50 mM (b) 100 mM (c) 150 mM (d) 200 mM solution of iron nitrate at 650$^{\circ}$C } \label{CONCENTRATION}
\end{center}
\end{figure}

\begin{figure}[h]
\begin{center}
\includegraphics[width=0.45\textwidth]{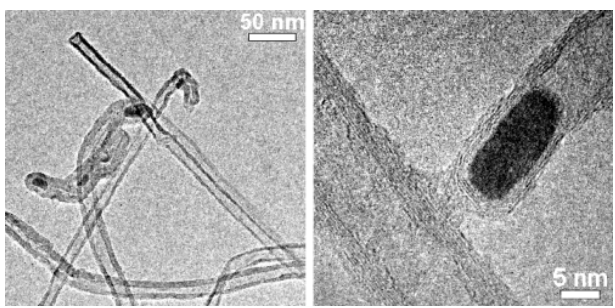} \caption {\it TEM micrograph of individual nanotubes grown at 650$^{\circ}$C with a 50~mM iron solution.} \label{TEM}
\end{center}
\end{figure}

Fig.~\ref{CONCENTRATION} shows clearly that the use of higher
catalyst concentrations resulted in an increase of the nanotube
density. The nanotubes reached lengths of several $\mu$m with a
diameter up to 25 nm. Transmission electron microscopy (TEM) of
the samples obtained at 650$^{\circ}$C using the 50~mM iron
catalyst confirmed that the structures are well-graphitized and
well-separated multiwalled nanotubes which are not filled
(Fig.~\ref{TEM}). Most of them have open ends and some nanotubes
contain encapsulated catalyst particles. These particles are
aligned in the growth direction and of prolate shape. They have
diameters of about 8~nm and lengths of about 16~nm.

\section{Discussion}

We suppose that the mechanism of the catalytic growth of carbon
nanotubes is similar to the one described by Kanzow et
al.~\cite{DING}. The acetylene is dissociated catalytically at
facets of well defined crystal orientation of a small metal
(oxide) particle. The resulting hydrogen H$_{2}$ is removed by the
gas flow whereas the carbon diffuses into the particle. For
unsaturated hydrocarbons this process is highly exothermic. When
the particle is saturated with carbon, the carbon segregates on
another, less reactive surface of the particle, which is an
endothermic process. The resulting temperature gradient supports
the diffusion of carbon through the particle. To avoid dangling
bonds, the carbon atoms assemble in a sp$^2$ structure at the
cooler side of the particle, which leads to the formation of a
nanotube.

The growth of nanotubes starts at temperatures of 620$^{\circ}$C
because there is enough mobility of the atoms of the catalyst to
enable the diffusion of the carbon through the particle and to
start the dynamic of the catalytic construction of carbon
nanotubes. At temperatures above 800$^{\circ}$C the acetylene
starts to dissociate already in the gas phase. The carbon in the
gas flow of acetylene and nitrogen forms carbon flakes which will
be adsorbed on the surface of the sample as well as on the surface
of the structures (see Ref.~\cite{ELSER}). Since the acetylene is
not completely dissociated and the catalytic growth proceeds, the
nanotubes will still grow and be covered with a layer of carbon
formed by the carbon flakes. The structures get thicker with
temperature because the proportion between dissociated and
molecular acetylene increases.

We found only few nanotubes using nickel and cobalt as catalysts,
whereas other groups successfully used
nickel~\cite{PROVENCIO,KIM2}, nickel-cobalt~\cite{KIM} or
cobalt~\cite{LUCAS}. This may be due to several facts. First, the
catalyst is usually deposited by thermal evaporation or sputtering
from a pure metal source. Furthermore, the catalyst is often
reduced before the growth or the growth itself is carried out in a
reducing atmosphere to ensure that the catalyst remains
metallic~\cite{RAO}. In our case however, the catalyst is
dissolved in a solution. Since the printed iron catalyst consists
of a gel-like material that forms a Fe$_{2}$O$_{3}$ film after
annealing~\cite{KERN} and no step is taken to reduce the catalyst,
we conclude that in our case the catalyst is not pure metal, but
metal oxide. This may significantly change the behavior of the
catalyst. Second, we use thermal CVD, in contrast to the Hot
Filament or Plasma Enhanced CVD used in other studies. The hot
filament and the plasma provides an additional possibility for the
dissociation of the hydrocarbon and may decisively influence the
reaction kinetics.

The electronic properties and the enormous length-diameter ratio
of the carbon nanotubes offers the possibility to use them as
field emitters~\cite{CHATELAIN}. The non-aligned arrangement of
the carbon nanotubes is known to be even more efficient in field
emission than an aligned one~\cite{KERN3}. The emission from
aligned nanotube films is lower because of screening effects
between densely packed neighboring tubes and the small height of
the few protruding tubes. In contrast, the non-aligned films offer
well separated nanotubes which do not show these
effects~\cite{CHATELAIN2}. For the sake of profitability their
application in field emission displays requires glass as substrate
instead of silicon, but the borosilicate glass used melts around
660$^{\circ}$C. Most studies carried out on catalytic nanotube
growth use however temperatures above
700$^{\circ}$C~\cite{LUCAS,KIM}, which are too high for that
purpose. Choi et al. report a plasma-enhanced CVD process at a
temperature of 550$^{\circ}$C \cite{KIM2}, but the diameter of
these nanotubes seems quite large and lacks uniformity. We could
demonstrate that the growth of thin carbon nanotubes with just a
few layers of carbon starts around 620$^{\circ}$C, and that high
quality films are obtained at 650$^{\circ}$C. These nanotubes have
uniform diameter and well-graphitized walls which is an indicator
for good field emission properties. The temperature may even be
further lowered by using other gas mixtures, by using other
metallic or heterogeneous catalysts~\cite{HAN}.

\section{Conclusions}

We observed a significant influence of the temperature and the
catalyst material on the quality of the carbon nanostructures. The
diameter of the nanotubes and the density is adjustable by
choosing the corresponding temperature and/or the concentration of
the catalyst solution. Under the studied conditions iron is the
best catalyst. We observed a morphology transition with
temperature from multi-wall nanotubes to ``carbon worms''. The
best nanotubes are obtained at temperatures between 650$^{\circ}$C
and 720$^{\circ}$C. Nanotubes obtained at temperatures below the
melting temperature of borosilicate glass of 660$^{\circ}$C are
suitable as field emitters for flat panel displays.

\section*{Acknowledgement}

The Swiss National Science Foundation (SNF) is acknowledged for
the financial support. The electron microscopy was performed at
the Centre Interd\'epartmental de Microscopie Electronique (CIME)
of EPFL.


\section*{Figures}

\clearpage

\end{document}